\documentclass[10pt,conference,prologue,dvipsnames]{IEEEtran}

\AtBeginDocument{%
  \providecommand\BibTeX{{%
    \normalfont B\kern-0.5em{\scshape i\kern-0.25em b}\kern-0.8em\TeX}}}

\usepackage{float}
\usepackage[dvipsnames]{xcolor}
\usepackage[utf8]{inputenc} 
\usepackage[T1]{fontenc}    
\usepackage{hyperref}       
\usepackage{url}            
\usepackage{booktabs}       
\usepackage{amsfonts}       
\usepackage{nicefrac}       
\usepackage{microtype}      
\usepackage{tcolorbox}
\usepackage{adjustbox}
\usepackage[frozencache,cachedir=.]{minted}
\usepackage{array}
\usepackage{siunitx}
\setminted{fontsize=\scriptsize}
\definecolor{bg}{HTML}{282828}
\usepackage{pmboxdraw}
\usepackage{multirow}
\usepackage[keeplastbox]{flushend}

\begin{document}

\newcommand{\ie}{\textit{i.e.,}~}
\newcommand{\eg}{\textit{e.g.,}~}
\newcommand{\etc}{\textit{etc.}~}
\newcommand{\etal}{\textit{et al.}~}

\newcommand{\nb}[2]{
    \fbox{\bfseries\sffamily\scriptsize#1}
    {\sf\small$\blacktriangleright$\textit{#2}$\blacktriangleleft$}
}

\newcommand\MICHELE[1]{\textcolor{blue}{\nb{MICHELE}{#1}}}
\newcommand\DAWN[1]{\textcolor{green}{\nb{DAWN}{#1}}}
\newcommand\ALEXEY[1]{\textcolor{red}{\nb{ALEXEY}{#1}}}

\title{Generating Accurate Assert Statements for Unit Test Cases using Pretrained Transformers}

\author{\IEEEauthorblockN{Michele Tufano, Dawn Drain, Alexey Svyatkovskiy, Neel Sundaresan}
\IEEEauthorblockA{Microsoft\\
Redmond, WA, USA\\
Email: \{mitufano, dadrain, alsvyatk, neels\}@microsoft.com}\vspace{-5ex}}

\maketitle

\begin{abstract}
Unit testing represents the foundational basis of the software testing pyramid, beneath integration and end-to-end testing. Automated software testing researchers have proposed a variety of techniques to assist developers in this time-consuming task.

In this paper we present an approach to support developers in writing unit test cases by generating accurate and useful assert statements. Our approach is based on a state-of-the-art transformer model initially pretrained on an English textual corpus. This semantically rich model is then trained in a semi-supervised fashion on a large corpus of source code. Finally, we finetune this model on the task of generating assert statements for unit tests.

The resulting model is able to generate accurate assert statements for a given method under test. In our empirical evaluation, the model was able to predict the exact assert statements written by developers in 62\% of the cases in the first attempt. The results show 80\% relative improvement for top-1 accuracy over the previous RNN-based approach in the literature. We also show the substantial impact of the pretraining process on the performances of our model, as well as comparing it with assert auto-completion task. Finally, we demonstrate how our approach can be used to augment EvoSuite test cases, with additional asserts leading to improved test coverage.
\end{abstract}


\section{Introduction}
Software testing is recognized as one of the most crucial, challenging, and expensive parts of the software lifecycle. It comes as no surprise that the software testing research community has invested significant effort in designing approaches that aim at supporting or automating software testing activities. An example of this endeavor is the work targeting the automatic generation of unit tests \cite{fraser2011evosuite, pacheco2007randoop}. While these works represent a notable achievement towards the goal of automated testing, they come with several limitations, recently highlighted by studies in industrial settings \cite{almasi2017industrial, shamshiri2015automated}. 

One of the major challenges these tools aim to overcome is the generation of accurate assert statements, which have been found to be often incomplete or inadequate to properly test the behavior of a software component. Generating meaningful assert statements is one of the key challenges in automatic test case generation \cite{watson2020learning}. Assert statements represent the basic blocks of software testing, used by developers to check conditions or states in a program and reason about program correctness. 

Watson \etal recently proposed ATLAS \cite{watson2020learning}, an RNN-based approach  which aims at learning from thousands of unit test methods how to generate meaningful assert statements. Inspired by this work, we improve upon it in substantial ways.

In this paper we present an approach to generate accurate assert statements based on state-of-the-art transformer model and relying on transfer learning to achieve best-in-class performances, predicting the correct assert in 62\% of the cases in the very first attempt, which represents an 80\% relative improvement for top-1 accuracy over the previous work \cite{watson2020learning}.

\begin{figure*}[ht]
    \centering
    \includegraphics[width=0.99\textwidth]{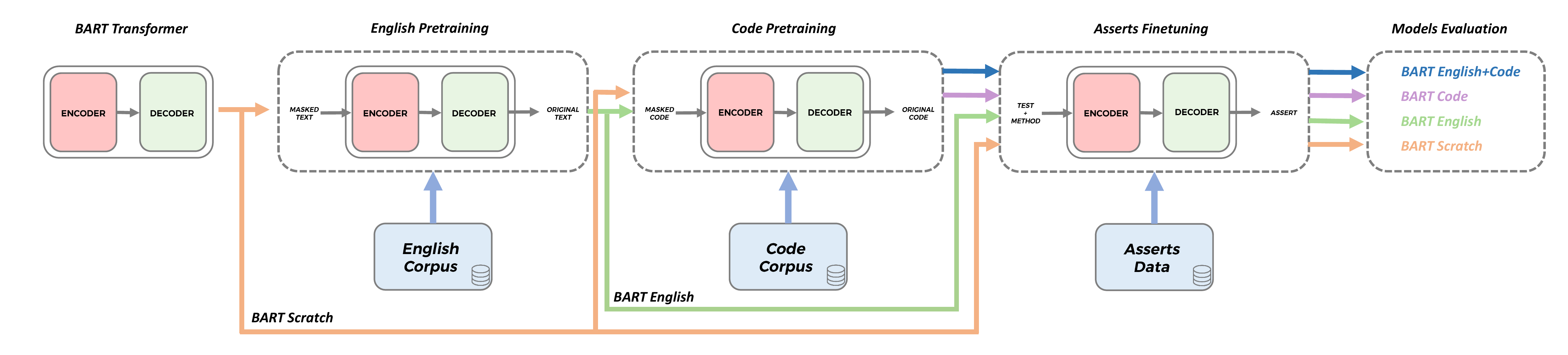}
    \caption{Overview of the Model Training Process. Starting from a BART Transformer model, we perform English and Source Code pretraining, and finetune the models on the Assert generation task. We obtain four different models based on the level of pretraining performed.}
    \label{fig:overview}
\end{figure*}

Transfer learning is a technique which first trains a model in an unsupervised fashion on large quantities of unlabeled data, and then finetunes on a downstream task like classification or translation. This technique has emerged as a standard route to achieve state-of-the-art results in natural language processing (NLP) tasks, obtaining higher performance and requiring much less resources than training each task from scratch~\cite{gpt2, bert, xlnet}. The intuitive explanation for this success is that a model which learns to generate text or fill in blanks will have developed biases reflected in the data which can help it learn to perform particular language tasks faster and with higher performance. 

In this paper, we extend this idea to source code as a language for a task of automated assert statement generation. We pretrain a sequence-to-sequence transformer model on a large source code and English language corpora, and finetune it on an assert statement generation task.

In our extensive empirical evaluation we assess several properties of our approach, such as intrinsic model metrics as well as extrinsic metrics related to the generated asserts. Finally, we evaluate the proposed approach in a scenario where it is used to support automated test case generation tools, such as EvoSuite, by augmenting the test cases with the assert statements generated by our approach.

To summarize, this paper provides the following contributions:
\begin{itemize}
    \item An approach that generates accurate assert statements based on a sequence-to-sequence transformer model. Our approach can predict the correct assert in 62\% of the cases in the very first attempt, and reaches up to 84\% correctness when allowing the model to suggest more asserts.
    
    \item We empirically demonstrate the benefits of pretraing on both English and source code corpora for the downstream assert generation task, resulting in performance improvement in terms of variety of intrinsic metrics such as BLEU score, validation loss, and syntactic correctness.
    
    \item We investigate how our proposed approach can be used to augment existing test cases, such as those generated by EvoSuite, with additional assert statements that lead to test coverage improvements.
\end{itemize}

\bigskip

\section{Approach}
Figure \ref{fig:overview} provides an overview of the training pipeline we followed in building our models specialized in assert statement generation.

We begin with the state-of-the-art BART Transformer (Sec. \ref{sec:transformer}) which will serve as the reference architecture for our models. We employ two pretraining stages: English Pretraining (Sec. \ref{sec:english_pretraining}) where we perform semi-supervised pretraining on a large corpus of English text; Code Pretraining (Sec. \ref{sec:code_pretraining}) where the model is pretrained on Java source code. Next, we perform the finetuning on the task of generating assert statements for unit test cases (Sec. \ref{sec:finetuning}), relying on a labelled dataset of test cases and method under tests. Finally, we evaluate variants of the models (Sec. \ref{sec:models}) obtained with different levels of pretraining indicated with different arrows in Figure \ref{fig:overview}.

\subsection{BART Transformer}
\label{sec:transformer}

BART~\cite{lewis2019bart} is a denoising autoencoder which utilizes the standard sequence-to-sequence transformer architecture from~\cite{DBLP:journals/corr/VaswaniSPUJGKP17}, substituting ReLUs with GeLU activation functions.

We select the BART model architecture because it facilitates finetuning for downstream translation task of assert statement generation, providing a more advanced set of noising transformations, which include token masking, token deletion, infilling and statement permutation. The model is pretrained by corrupting documents and optimizing the cross-entropy loss between the decoder’s output and the original input sequence. 

We pretrain the BART large model architecture, having 12 layers in the encoder and decoder. The model is trained in mixed-precision, using Adam stochastic optimization procedure with $\epsilon=10^{-6}$, and $\beta_1=0.9$, $\beta_2=0.98$ optimizer parameters; we apply inverse square root learning rate schedule with the base learning rate of 0.0001, a warmup period of 5000 update steps, and the local gradient accumulation with a frequency of 4 update steps.

\subsection{English Pretraining}
\label{sec:english_pretraining}
In this stage we pretrain a model in a semi-supervised fashion on a large corpus of English text, with the goal of learning semantic and statistical properties of natural language.

\subsubsection{Dataset}
The pretraining is performed for 40 epochs on 160GB of English text extracted from books, Wikipedia, and news articles~\cite{liu2019roberta}, comprising a total of X lines of text.

\subsubsection{Training Strategy}
BART is trained in an unsupervised manner. Given corrupted text, its objective is to reconstruct the original text. The particular type of noise used in this work involves masking 30\% of all tokens, with masks covering spans of tokens with lengths following a Poisson distribution parameterized by $\lambda=3$, as well as permuting all sentences.

\subsection{Code Pretraining}
\label{sec:code_pretraining}
In this stage we pretrain a model on source code corpus written in Java language, with the goal of learning syntax and properties of source code.

\subsubsection{Dataset}
We collect this code corpus dataset by crawling all public, non-fork Java repositories on GitHub with at least 50 stars. We then deduplicate at the file-level using a hash function. After filtering based for permissive licenses and filtering out based on heuristics like the fraction of non-ascii characters, we are left with 25GB of training data from the 26,000 repos. For pretraining validation, we use the 239 test Java repos from CodeSearchNet \cite{husain2019codesearchnet}, which comprise 600MB. 

\subsubsection{Training Strategy}
A similar pretraining strategy to English pretraining is employed. The source code files are corrupted by deleting 20\% of all tokens independently and rotating half of all documents. This pretraining is performed for 10 epochs.

\subsection{Asserts Finetuning}
\label{sec:finetuning}
In this stage we finetune a model on the task of generating assert statements for unit test cases. Specifically, we represent this task as a \textit{translation} task, where the source is the partially written unit test along with the method under test, and the target is the correct assert statement that the developer wrote for that unit test.

\subsubsection{Dataset}
To perform the finetuning, we rely on the publicly available dataset of unit test methods used to evaluate ATLAS \cite{watson2020learning}. This dataset is comprised of Test Methods (\ie methods within a unit test case), Focal Methods (\ie the methods under test), and Asserts (\ie the assert statements within the Test Methods).

This dataset has been mined from more than 9 thousand open-source GitHub projects containing unit test cases defined with JUnit. The authors first extract methods beginning with the \texttt{@Test} annotation as candidate Test Methods. From these candidate methods the authors select those that specify a single assert statement. Next, for each Test Method they pinpoint the Focal Method (\ie the method that the current Test Method is testing) using a heuristic \cite{qusef2010recovering} which looks at last method invocation before (or within) the assert statement. Finally, each Test Method is modified by replacing the Assert statement with a placeholder \texttt{<AssertPlaceholder>}.

Each data point in the dataset is referred to as a Test-Assert Pair (TAP), and can be seen as a triplet $TAP_i=\{tm'_i, fm_i, a_i\}$ where $tm'_i$ is the Test Method where the assert statement has been replaced with a placeholder, $fm_i$ is the Focal Method, and $a_i$ is the assert statement. This data is organized as a parallel corpus, a common format for machine translation tasks, where the source sentence $s_i = \{ tm'_i + fm_i\}$ is the concatenation of the Test Method and Focal Method, while the target sentence $t_i = a_i$ is the assert statement to predict. Figure \ref{fig:assert_data} provides an example of a TAP. The test method $tm_i$ is \texttt{testLength()} and its corresponding focal method $fm_i$ is \texttt{length()}. The test method creates two sets of bits and check that their length is the same, using the \texttt{assertEquals} statement. The source sentence is the concatenation of the test method and focal method, where the assert statement is replaced with a placeholder. The corresponding target output is the assert to be predicted.

For our models, we use the \textit{raw} version of the dataset -- corpus comprising the original source code tokens -- rather than the \textit{abstract} version (where some tokens are replaced with IDs), since we aim to exploit the rich semantics of all the variable and method names. We keep the original split of the dataset in training, validation, and testing sets (80\%, 10\%, 10\%) for a fair comparison. Table \ref{tab:datasets} reports the number of instances in the dataset.

\subsubsection{Training Strategy}
The finetuning process is a translation task, where we train the model to learn the mapping $s_i \rightarrow t_i$ as a conditional probability $P(a_i \vert  tm'_i + fm_i)$.

During training we use the cross entropy loss and the Adam optimizer and monitor the loss on the validation set for early stopping.

We use shared vocabulary embeddings between Encoder and Decoder for optimizations reasons \cite{DBLP:journals/corr/VaswaniSPUJGKP17, DBLP:journals/corr/PressW16} and because our input and output language is the same (\ie Java source code).

\subsection{Model Variants}
\label{sec:models}
At the end of these stages, we obtain four different variants of the model, based on the level of pretraining performed:
\begin{itemize}
    \item \textit{BART\_Scratch}: A model which has not been pretrained on any corpus but directly finetuned on the assert generation task. This model represents the orange line in Figure \ref{fig:overview}.
    \item \textit{BART\_English}: A model which has been pretrained on the English corpus and then finetuned for the assert generation task. This model represents the green line in Figure \ref{fig:overview}.
    \item \textit{BART\_Code}: A model pretrained on the source code corpus, then finetuned on the assert generation task. This model represents the purple line in Figure \ref{fig:overview}.
    \item \textit{BART\_English+Code}: A model pretrained first on English and further pretrained on source code corpus, then finetuned on the assert generation task. This model represents the blue line in Figure \ref{fig:overview}.
\end{itemize}

\begin{figure}[t]
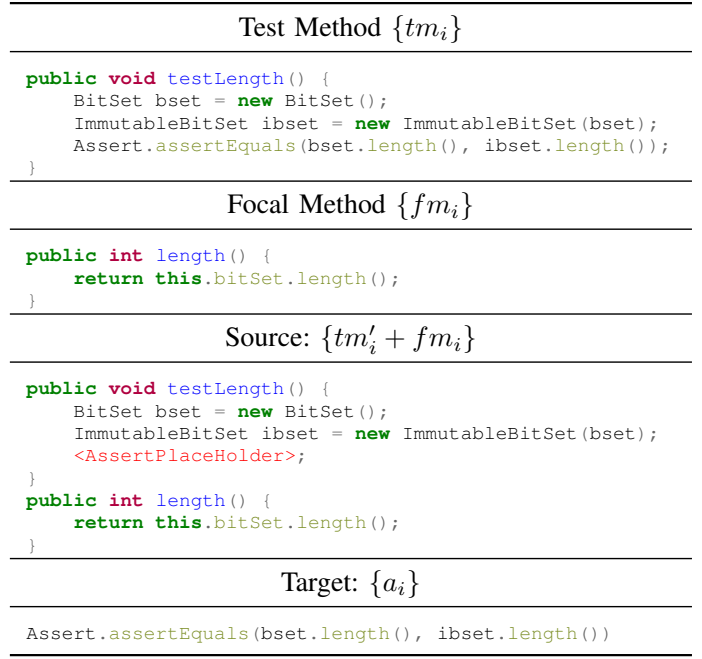

\vspace{-0.0cm}
    \centering
\begin{adjustbox}{width=0.5\textwidth}
\begin{tabular}{c}
\toprule
  Test Method $\{tm_i\}$ \\
\midrule
\begin{minipage}[t]{0.45\textwidth}
\begin{minted}{java}
public void testLength() { 
    BitSet bset = new BitSet(); 
    ImmutableBitSet ibset = new ImmutableBitSet(bset); 
    Assert.assertEquals(bset.length(), ibset.length());
}
\end{minted}
\end{minipage}\\
\midrule
Focal Method $\{fm_i\}$ \\
\midrule
\begin{minipage}[t]{0.45\textwidth}
\begin{minted}{java}
public int length() { 
    return this.bitSet.length(); 
}
\end{minted}
\end{minipage} \\
\midrule
Source: $\{ tm'_i + fm_i\}$\\
\midrule
\begin{minipage}[t]{0.45\textwidth}
\begin{minted}[escapeinside=||]{java}
public void testLength() { 
    BitSet bset = new BitSet(); 
    ImmutableBitSet ibset = new ImmutableBitSet(bset); 
    |\color{red}{<AssertPlaceHolder>}|;
}
public int length() { 
    return this.bitSet.length(); 
}
\end{minted}
\end{minipage} \\
\midrule
Target: $\{a_i \}$\\
\midrule
\begin{minipage}[t]{0.45\textwidth}
\begin{minted}{java}
Assert.assertEquals(bset.length(), ibset.length())
\end{minted}
\end{minipage}
\\
\bottomrule
\end{tabular}
\end{adjustbox}
\vspace{-0.0cm}
\caption{Example of a Test-Assert Pair (TAP)\protect\\The source is formed by concatenating the partial test case (without assert) and the focal method. The target is the assert statement to generate.}
\vspace{-0.0cm}
\label{fig:assert_data}
\end{figure}

\section{Experimental Design}
The goal of our empirical study is to determine if our approach can generate accurate assert statements for unit test cases, in one or very few attempts. We investigate whether our approach outperforms the previous RNN-based approach ATLAS \cite{watson2020learning}. Additionally we explore the impact of different pretrainings on the assert generation performances as well as the effect of incorporating the focal method in the input to the model.

Our experiments aims at answering the research questions described in the following paragraphs. \smallskip

\textbf{RQ$_1$: How does our approach compare with ATLAS?}

We compare the performances of our models against ATLAS \cite{watson2020learning}, the RNN-based approach available in the literature. Specifically, to ensure fair comparison, we perform the finetuning process of our models on the exact same training set, and evaluate and compare the performances on the same test set. To compare the models' performances we use the top-k accuracy metric, which measures the accuracy of a model with different number of attempted predictions. In particular, if the target assert statement $a_i$ for the \textit{i}-th input in the test set, is in the top-k predictions of the model, we count it as a correct prediction at $k$. Similarly to ATLAS \cite{watson2020learning}, we experiment with $k=\{1, \dots, 50\}$ with a maximum beam size of 50.
\smallskip

\textbf{RQ$_2$: What is the effect of the pretraining process on the assert generation task?}

We investigate whether pretraining our models on English corpus, on source code, and both corpora has any noticeable impact on the downstream performances of the models on the assert generation task. In particular, we compare the \textit{BART\_Scratch} model, which has not been pretrained on any corpus, against \textit{BART\_English} which was pretrained on English corpus, \textit{BART\_Code} which was pretrained on source code, and \textit{BART\_English+Code} which was pretrained on both English and source code. This comparison is performed considering:
\begin{itemize}
    \item \textit{Extrinsic metrics}: the top-k accuracy on the assert generation task;
    \item \textit{Intrinsic metrics}: (i) best validation loss and the number of training steps required to reach it (\eg faster convergence); (ii) BLEU4 score, a common metric for machine translation tasks, evaluated on the test set; (iii) the syntactic correctness of the generated asserts, determined using a Java Parser.
\end{itemize}

\smallskip

\textbf{RQ$_3$: What is the effect of the Focal Method on the performance of the model?}

In Section \ref{sec:finetuning} we described the finetuning process, where the input to the Encoder is the concatenation $\{ tm'_i + fm_i\}$ of the Test Method $tm'_i$ and Focal Method $fm_i$. In this research question our goal is to understand the effect on performances of the Focal Method as input to the model, when generating the assert statements. In particular, we select our best model obtained from RQ$_1$ and RQ$_2$ and compare against an equivalent model (\ie same preprocessing steps) but finetuned on a parallel corpus that does not contain the focal method, trying to learn the probability $P(a_i \vert  tm'_i)$. Specifically, this can be seen as an auto-completion task, where the source is the partially written test method $tm'_i$ and the expected target output is the assert $a_i$. We compare the performances of the models with or without the focal method in terms of top-k accuracy on the test set.

\smallskip

\textbf{RQ$_4$: What is the quality of the generated asserts?}

In the last research question we investigate the quality of the assert statements generated by the model. We manually analyze instances of correct predictions as well as inspecting those that do not match with the target assert statement. We report qualitative examples and discussion.

\begin{table}[]
    \centering
	\vspace{-0.0cm}
	\caption{Datasets used for English and Source Code pretraining,\protect\\ and Assert finetuning}
	\label{tab:datasets}
	\resizebox{0.8\linewidth}{!}{
\begin{tabular}{lrrr}
\toprule
{Set} &  English &  Source Code &   Asserts \\
\midrule
Train & 160GB & 25GB & 150,523 \\
Valid & - & 600MB & 18,816 \\
Test & - & - & 18,815 \\
\midrule
Total & 160GB & 25GB & 188,154 \\
\bottomrule
\end{tabular}}
\vspace{-0.0cm}
\end{table}

\textbf{RQ$_5$: Can our approach be used to improve automatically generated test cases?}

The goal of this research question is to provide a preliminary investigation on the potential benefits of using our approach to improve automated test case generation tools, such as EvoSuite. Specifically, we aim at enhancing test cases generated by EvoSuite by inserting additional asserts generated by our approach. We evaluate the potential benefits in terms of test coverage boost and qualitative discuss the additional asserts.

For this investigation we select a small but reproducible testbed using defects4j. We rely on defects4j since it provides a reliable infrastructure to generate, compile, execute, and evaluate test cases. Specifically, we select Lang-1-f, which represents the fixed version of the first bug in the defects4j collection belonging to the project Apache Commons Lang.

To generate test cases with EvoSuite, we use the defects4j built-in command \texttt{gen\_tests.pl -g evosuite -p Lang -v 1f}. This command invokes EvoSuite test generation on the first fixed revision of Lang, which will generate test cases for the class affected by the bug (\ie \texttt{NumberUtils}). We let EvoSuite generate test cases for 500 seconds ($\sim8$ minutes) and compute the test coverage using defects4j which relies on Cobertura, singularly for each test case.

Next, we select the 18 unique focal methods of the class, without considering overloaded copies of the methods, and the corresponding test cases generated by EvoSuite. We select a single best test case for each of the focal methods. Once we have the mapped test case pair, we generate additional assert statements for each of the pair using our approach. Specifically, for each focal method we generate the top-10 predictions and select a single assert from them, which we insert as the last statement within the EvoSuite test case. Finally, we execute the newly augmented test cases and recompute the test coverage for each of them.

\section{Experimental Results}
In this section we report and discuss the results of our empirical study.

\begin{figure}
    \centering
    \includegraphics[width=0.49\textwidth]{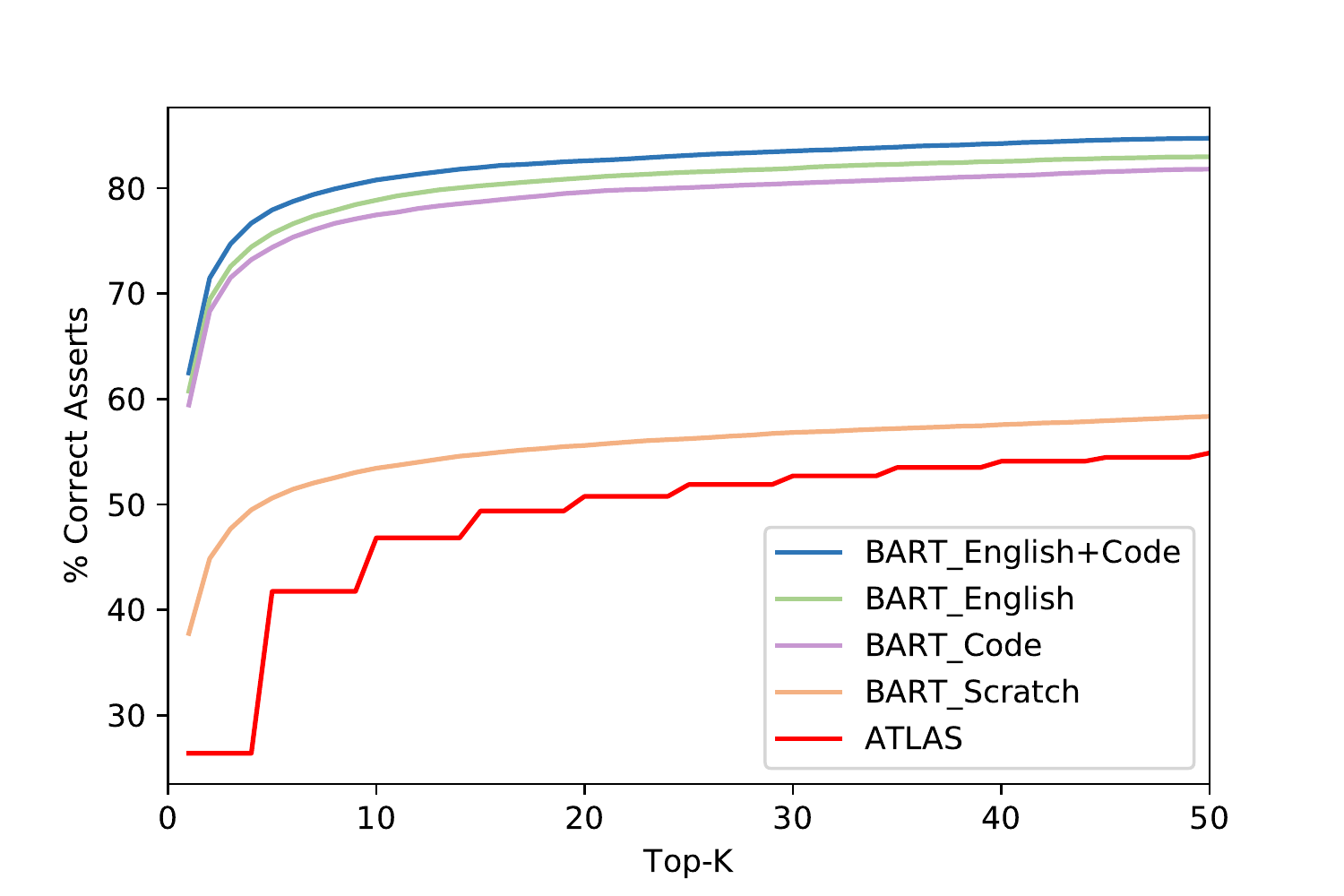}
    \vspace{-0.2cm}
    \caption{Top-K Accuracy Results. \protect\\Comparing our four BART model variants against ATLAS}
    \vspace{0.2cm}
    \label{fig:top-k}
\end{figure}

\textbf{RQ$_1$: How does our approach compare with ATLAS?}

Figure \ref{fig:top-k} reports the top-k accuracy for our four variations of the model as well as for ATLAS. The x-axis represents the $k$ value, ranging from 1 to 50, while the y-axis indicates the percentage of correct asserts in the test set. For ATLAS we report the results as they appear in the original work \cite{watson2020learning}, by considering the best model (\ie Abstract Model). It is worth noting that the ATLAS line is shaped as a step-function because the authors reports only the value of $k$ with 5 step increment (\eg $1, 5, 10, \dots$), while we computed the top-k accuracy at each integer value $k$ from 1 to 50, hence the smoother curve.

The results show that our models outperform ATLAS at any $k$ value. In particular, \textit{BART\_English+Code} can correctly predict the target assert statement (originally written by the developer) in 62\% of the cases in just the first attempt. This represents an $80\%$ relative improvement over the top-1 ATLAS accuracy of 27\%. 

The impressive results on the top-1 accuracy are particularly important in terms of usability and applicability of this approach beyond research, in actual development environments. Practically, developers would obtain accurate and relevant assert statements in one or very few suggestions, without the need of going through a long list while discarding incorrect recommendations.

\vspace{-0.0cm}
\begin{center} 
\fbox{
\begin{minipage}[t]{0.97\linewidth}
{\bf Summary for RQ$_1$.} 
Our approach outperforms ATLAS with a relative improvement of 80\% on top-1 accuracy.
\end{minipage}
}
\end{center}

\smallskip

\textbf{RQ$_2$: What is the effect of the pretraining process on the assert generation task?}
\subsubsection*{Extrinsic Metrics}
Figure \ref{fig:top-k} shows a massive gap between the performance of the model whiteout pretraining (\textit{BART\_Scratch}) compared to the models with English (\textit{BART\_English}), source code (\textit{BART\_Code}), and both (\textit{BART\_English+Code}) pretraining. Specifically, the performance gap between \textit{BART\_Scratch} and the models with single pretraining phase (\textit{BART\_Code} or \textit{BART\_English}) is 22-25\%,  while the gap between \textit{BART\_English} and \textit{BART\_English+Code} is 1.54-2.25\%, in favor of the model which was pretrained on both English and source code.

These results highlight the importance of pretraining on the performance of downstream tasks. It is particularly striking the effect of pretraining on natural language English text over a downstream task involving source code. This result emphasizes the significance of having a model which understands the semantics of variable and method names in the code.

While additional pretraining on source code appears to have a limited impact on performances, compared to pretraining only on English, it still delivers consistent improvements, which could potentially be higher on bigger test sets.

\subsubsection*{Intrinsic Metrics}
\begin{figure}
    \centering
    \includegraphics[width=0.49\textwidth]{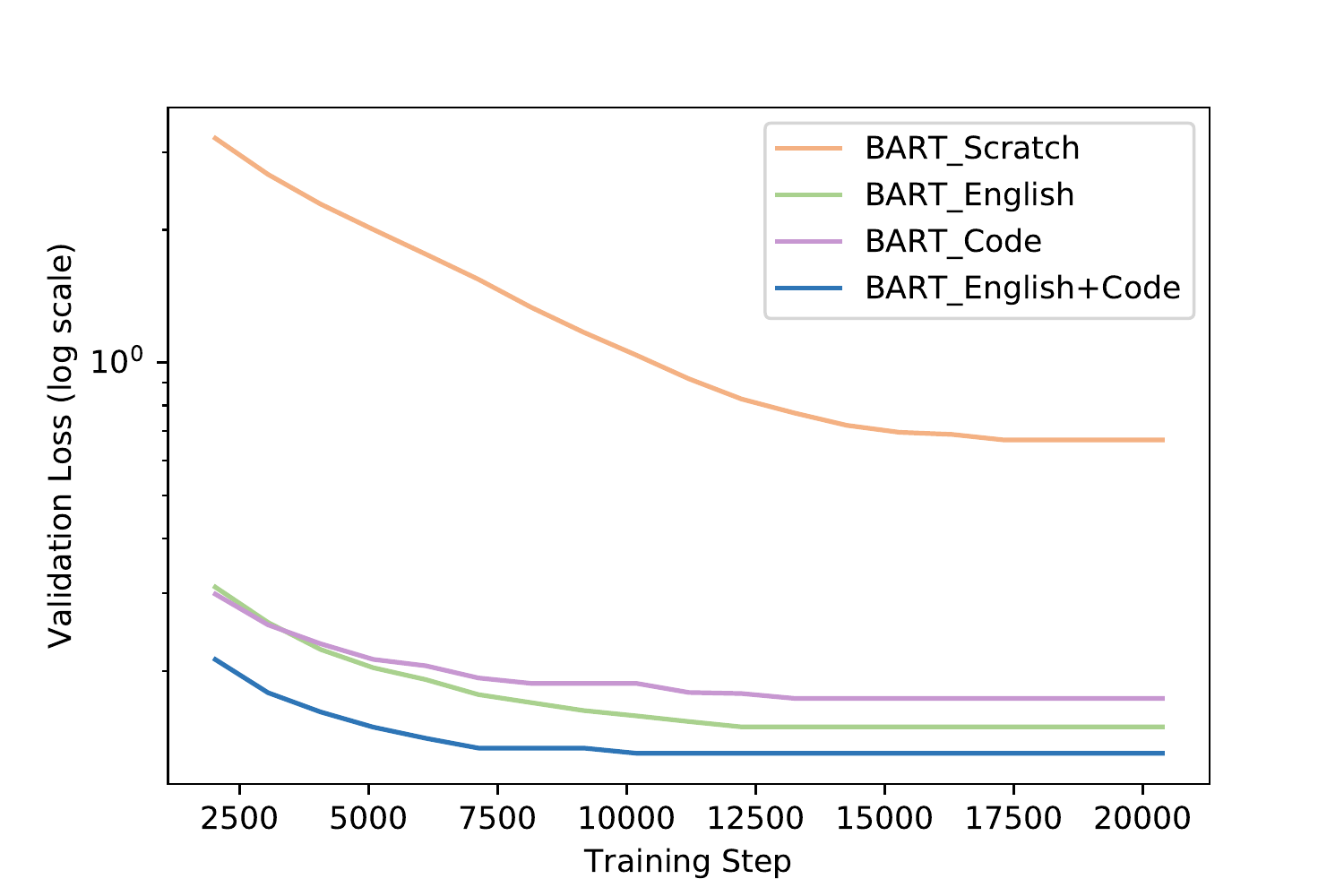}
    \caption{Validation Loss obtained during Assert finetuning for our four BART model variants}
    \vspace{-0.0cm}
    \label{fig:loss}
\end{figure}

In terms of intrinsic metrics, Figure \ref{fig:loss} shows the cross-entropy loss on the validation set during training for the four model variations. Similarly to what observed with the extrinsic metric, we note a substantial gap between the model without pretraining (\textit{BART\_Scratch}) compared to the two models with English (\textit{BART\_English}), source code (\textit{BART\_Code}) and both (\textit{BART\_English+Code}) pretraining. Comparing the English only and the English+Code models, the additional pretraining on source code has three evident effects: (i) lower initial loss (0.21 vs 0.31); (ii) lower best loss (0.13 vs 0.15); (iii) faster convergence ($\sim$2500 training steps earlier).

Table \ref{tab:metrics} reports the intrinsic metrics computed for the four model variations. Specifically, \textit{BART\_English+Code} obtains the best BLEU4 and validation loss. Regarding the syntactic correctness, the model pretrained on both English and source code obtains the best value for the top single prediction, however when computing the correctness considering the top 25, and 50 generated asserts for each input in the test set, the model pretrained exclusively on source code achieves the highest correctness score. This result is somewhat predictable, since \textit{BART\_Code} has been pretrained and finetuned exclusively on source code, thus it should have the most consistent results in terms of syntax.

Overall, we observe a significant positive effect of pretraining on English and source code for both extrinsic and intrinsic metrics. While additional pretraining on source code appears to have a smaller impact than English pretraining alone, it is worth noting that we observe consistent improvements across all the analyzed metrics, corroborating the beneficial effect of the source code pretraining. Additionally, the small gap could be due to the nature of the downstream task, where the output is a single-line assert statement, which could closely resemble a natural language sentence.

\begin{table}[t]
	\vspace{-0.0cm}
	\caption{Accurate Predictions}
	\vspace{-0.0cm}
	\label{tab:perfect_pred}
	\resizebox{\linewidth}{!}{
\begin{tabular}{lrrrrr}
\toprule
{Top-K} &  ATLAS &  \textit{BART\_Scratch }&   \textit{BART\_Code} & \textit{BART\_English} &   \textit{BART\_English+Code} \\
\midrule
1  &   4968         (26.40\%) &       7106               (37.77\%) &        11183 (59.44\%) &   11430               (60.75\%) &     11754	(62.47\%) \\
5  &   7857        (41.76\%) &          9522               (50.61\%) &      13996 (74.39\%) &   14244               (75.71\%) &      14665	(77.94\%) \\
10 &   8812        (46.83\%) &         10055               (53.44\%) &      14576 (77.47\%) &   14839               (78.87\%) &      15200	(80.79\%) \\
15 &   9291        (49.38\%) &         10304               (54.76\%) &      14811 (78.72\%) &   15096               (80.23\%) &      15423	(81.97\%) \\
20 &   9554        (50.78\%) &         10461               (55.60\%) &      14982 (79.63\%) &   15238               (80.99\%) &      15541	(82.60\%) \\
25 &   9764        (51.89\%) &         10582               (56.24\%) &      15063 (80.06\%) &   15339               (81.53\%) &      15639	(83.12\%) \\
30 &   9918        (52.71\%) &         10693               (56.83\%) &      15140 (80.47\%) &   15407               (81.89\%) &      15716	(83.53\%) \\
35 &  10068        (53.51\%) &         10763               (57.20\%) &      15207 (80.82\%) &   15478               (82.26\%) &      15786	(83.90\%) \\
40 &  10179        (54.10\%) &         10833               (57.58\%) &      15274 (81.18\%) &   15530               (82.54\%) &      15849	(84.24\%) \\
45 &  10247        (54.46\%) &         10903               (57.95\%) &      15347 (81.57\%) &   15586               (82.84\%) &      15912	(84.57\%) \\
50 &  10327        (54.89\%) &         10979               (58.35\%) &      15392 (81.81\%) &   15615               (82.99\%) &      15944	(84.74\%) \\
\bottomrule
\end{tabular}}
\vspace{-0.0cm}
\end{table}

\begin{table}[]
	\vspace{0.7cm}
	\caption{Intrinsic Evaluation Metrics}
	\label{tab:metrics}
	\resizebox{1\linewidth}{!}{
\begin{tabular}{lrrrr}
\toprule
{Metric} & \textit{BART\_Scratch} & \textit{BART\_Code} & \textit{BART\_English}  & \textit{BART\_English+Code}\\
\midrule
BLEU4 & 72.40 & 84.24 & 84.13 & \textbf{85.35} \\
\midrule
Validation Loss & 0.67 & 0.17 & 0.15 & \textbf{0.13}\\
\midrule
Syntax Top-1 & 99.54\% & 99.56\% & 99.57\% & \textbf{99.58\%} \\
Syntax Top-25 & 92.97\% & \textbf{94.05\%} & 93.01\% & 93.96\%\\
Syntax Top-50 & 85.05\% & \textbf{88.12\%} & 87.07\% & 87.26\%\\
\bottomrule
\end{tabular}}
\end{table}

\vspace{-0.2cm}
\begin{center} 
\fbox{
\begin{minipage}[t]{0.97\linewidth}
{\bf Summary for RQ$_2$.} 
Pretraining has a significant positive effect on the downstream performances. Pretrainig on English text boosts the performances of 23-25\%, while further pretraining on source code can yield additional $\sim$2\% improvement.
\end{minipage}
}
\end{center}

\newpage

\textbf{RQ$_3$: What is the effect of the Focal Method on the performances of the model?}

\begin{figure}[t]
    \centering
    \vspace{-0.0cm}
    \includegraphics[width=0.49\textwidth]{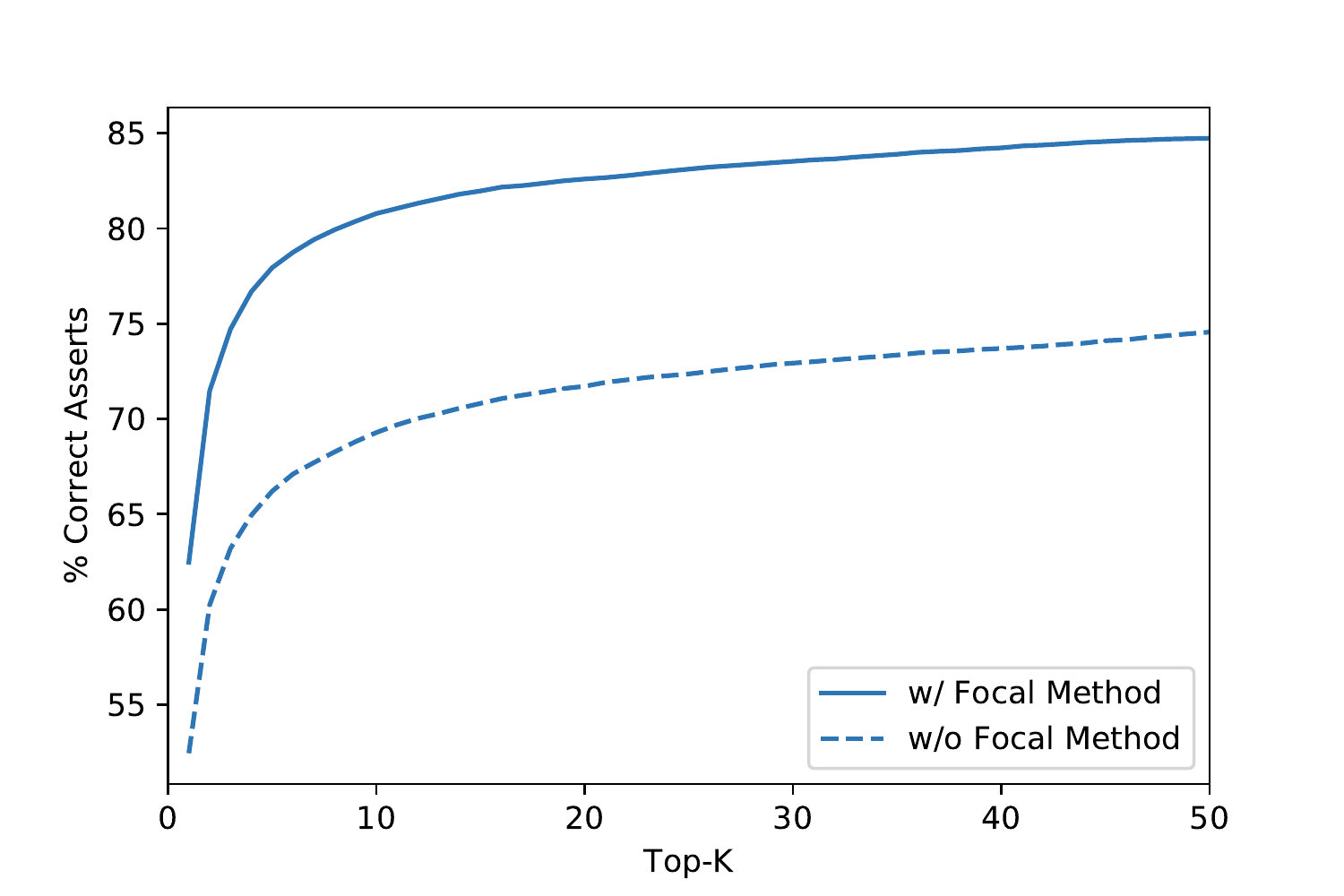}
    \vspace{-0.4cm}
    \caption{Top-K Accuracy - Comparing our model trained with or without the Focal Method as input}
    \vspace{-0.0cm}
    \label{fig:focal}
\end{figure}

To answer this research question, we compared the model that achieves the best performances in RQ$_1$ and RQ$_2$ -- \textit{BART\_English+Code} -- against a similar model (\ie same pretraining phases) but with different finetuning. Specifically, we selected the same model checkpoint after the source code pretraining, and fintuned the model on a modified dataset without the Focal Method as input. Figure \ref{fig:focal} shows the top-k accuracy of the models with (solid line) and without (dashed line) the Focal Method. The results show that the model which takes as input the Focal Method is more accurate in generating assert statements in $\sim$10\% of the cases. That is, there are certain assert statement that are not covered by the model w/o Focal Method, even when 50 different predictions are generated.
This result highlight the essential information provided by the Focal Method to inform the model on generating specific assert statements. 

\vspace{-0.0cm}
\begin{center} 
\fbox{
\begin{minipage}[t]{0.97\linewidth}
{\bf Summary for RQ$_3$.} 
The Focal Method provides essential information which allows the model to generate $\sim$10\% more accurate asserts compared to a generic auto-completion model.
\end{minipage}
}
\end{center}

\textbf{RQ$_4$: What is the quality of the generated asserts?}
To answer this research question we analyze and discuss examples of generated assert statements. Figure \ref{fig:genearted_asserts} provides examples of common, complex, and equivalent assert statements generated by our best model \textit{BART\_English+Code}. The list of common asserts comprises statements that are correctly predicted by the mode (\ie match the original assert) which are often found in test cases in different contexts. For example, these asserts usually check that the \texttt{result} is equal to the \texttt{expected} value, or that a given \texttt{list} contains an \texttt{element} that was previously added. These types of asserts are usually predicted in the very first attempts of the model. While these represent simple assert statements, they still require the model to detect the variables used within the test/focal method and their relationship.

The list of complex assert statements showcase some of the challenging asserts correctly predicted by the model. These asserts involve multiple method calls, parameters, attributes, and variables that are less common.

The list of equivalent assert statements show generated asserts that do not exactly match with the target assert (\ie these are not counted as correct asserts in the top-k accuracy), but they are equivalent to the developer's assert. For example, the model suggests to get the class literal directly with \texttt{AbstractService.class}, while the developer uses the method \texttt{getClass()} which, in turn, uses the same class literal. In another instance, the developer checks that \texttt{status == 0} is true, while the model suggests an equivalent check with \texttt{assertEquals(0, status)}. Similarly, the model suggests to use \texttt{assertSame} on two objects, rather than the \texttt{==} equivalence. Note that for all these cases, the model was eventually able to predict the correct assert (\ie perfect match) in the subsequent attempts.

Finally, Figure \ref{fig:complete_examples} reports two complete examples with source and target, correctly predicted by the model. In the first example, the generated assert checks that the \texttt{event} object created with the \texttt{eventFactory} is of the correct class instance. In the second example, the model generates a complex assert statement involving numerical literals and variables previously used to set-up the testing environment. Additionally, the example in Figure \ref{fig:assert_data} described in Sec. \ref{sec:finetuning} was also correctly predicted by the model in the very first attempt.

These results highlight the need for additional metrics beyond simple accuracy. In particular, metrics that can recognize and discern cases where the generated assert statement is different yet equivalent to the one created by human developers, as well as non-equivalent asserts that are also correct in that context. Additionally, there could be many locations in the code where the developers did not introduce assert statements but the model could suggest reasonable ones, which are currently not uncovered in the quantitative metrics. The main goal of this research question was precisely to fill this gap with a qualitative and manual analysis.

\begin{center} 
\fbox{
\begin{minipage}[t]{0.97\linewidth}
{\bf Summary for RQ$_4$.} 
Our models can generate common assert statements as well as complex ones involving method calls, parameters, and unusual variables. In several cases, the model generates equivalent assert statements to the developer's assert. 
\end{minipage}
}
\end{center}

\begin{figure}[t]
    \centering
\begin{adjustbox}{width=0.5\textwidth}
\footnotesize
\begin{tabular}{c}
\toprule
Common Assert Statements\\
\midrule
\begin{minipage}[t]{0.45\textwidth}
\begin{minted}{java}
assertEquals(expected, result)
assertSame(rows, actual)
assertNotNull(client)
assertTrue(list.contains(element))
assertArrayEquals(expected, values)
assertEquals(1, result.getSize())
\end{minted}
\end{minipage} \\
\midrule
Complex Assert Statements\\
\midrule
\begin{minipage}[t]{0.45\textwidth}
\begin{minted}[fontsize=\tiny]{java}
assertEquals(0, zero.getPartialDerivative(n), epsilon [ n ])
assertThat(emptySession.getEnd(), CoreMatchers.equalTo(date))
assertEquals(container.getSoundEffects().read(0), Sound.ENTITY_CAT_HISS)
\end{minted}
\end{minipage} \\
\midrule
Equivalent Assert Statements\\
\midrule
\begin{minipage}[t]{0.45\textwidth}
\begin{minted}[fontsize=\tiny,escapeinside=||]{java}
|\textcolor{ForestGreen}{target}|:     assertNull(sm.get(serviceStub.getClass()))
|\textcolor{RoyalBlue}{predicted}|:  assertNull(sm.get(AbstractService.class))
\end{minted}
\end{minipage} \\
\midrule
\begin{minipage}[t]{0.45\textwidth}
\begin{minted}[fontsize=\tiny,escapeinside=||]{java}
|\textcolor{ForestGreen}{target}|:     assertTrue( status == 0)
|\textcolor{RoyalBlue}{predicted}|:  assertEquals(0, status)
\end{minted}
\end{minipage} \\
\midrule
\begin{minipage}[t]{0.45\textwidth}
\begin{minted}[fontsize=\tiny, escapeinside=||]{java}
|\textcolor{ForestGreen}{target}|:     assertEquals(user.getSNetVisibility(), visibility)
|\textcolor{RoyalBlue}{predicted}|:  assertEquals(visibility, user.getSNetVisibility())
\end{minted}
\end{minipage} \\
\midrule
\begin{minipage}[t]{0.45\textwidth}
\begin{minted}[fontsize=\tiny,escapeinside=||]{java}
|\textcolor{ForestGreen}{target}|:     assertTrue( ps1 == ps2)
|\textcolor{RoyalBlue}{predicted}|:  assertSame(ps1, ps2)
\end{minted}
\end{minipage} \\
\bottomrule
\end{tabular}
\end{adjustbox}
\vspace{0.1cm}
\caption{Examples of generated asserts\protect\\Common assert statements found in different contexts\protect\\Complex assert statement involving multiple method calls and parameters\protect\\Equivalent assert statements to the original target statement}
\label{fig:genearted_asserts}
\end{figure}

\begin{figure}[h]
    \centering
\begin{adjustbox}{width=0.5\textwidth}
\begin{tabular}{c}
\toprule
Source: $\{ tm'_i + fm_i\}$\\
\midrule
\begin{minipage}[t]{0.45\textwidth}
\begin{minted}[fontsize=\tiny, escapeinside=||]{java}
public void createBeginNwhinInvocation() { 
    Event event = eventFactory.createBeginNwhinInvocation(); 
    |\color{red}{<AssertPlaceHolder>}|;
}
public BeginNwhinInvocationEvent createBeginNwhinInvocation() { 
    return new BeginNwhinInvocationEvent(); 
}
\end{minted}
\end{minipage} \\
\midrule
Target: $\{a_i \}$\\
\midrule
\begin{minipage}[t]{0.45\textwidth}
\begin{minted}[fontsize=\tiny]{java}
Assert.assertTrue( event instanceof BeginNwhinInvocationEvent)
\end{minted}
\end{minipage}
\\
\midrule
Source: $\{ tm'_i + fm_i\}$\\
\midrule
\begin{minipage}[t]{0.45\textwidth}
\begin{minted}[fontsize=\tiny, escapeinside=||]{java}
public void simpleInsertTest() { 
    LRU lru = LRU(5, true); 
    for(int i = 0; i < 5; i ++) { 
        addAndExpectNoEviction(lru,(100 + i)); 
    } 
    
    for(int i = 5; i < 10; i ++) { 
        addAndExpectEviction(lru,(100 + i),(( 100 + i) - 5)); 
    } 
    for(int i = 5; i < 10; i ++) { 
        |\color{red}{<AssertPlaceHolder>}|; 
    } 
}
public boolean exists(long) { return m_lruMap.containsKey(id); }
\end{minted}
\end{minipage} \\
\midrule
Target: $\{a_i \}$\\
\midrule
\begin{minipage}[t]{0.45\textwidth}
\begin{minted}[fontsize=\tiny]{java}
Assert.assertTrue(lru.exists( 100 + i))
\end{minted}
\end{minipage}
\\
\bottomrule
\end{tabular}
\end{adjustbox}
\vspace{0.1cm}
\caption{Two Complete Examples of perfect predictions}
\label{fig:complete_examples}
\end{figure}

\begin{table*}[t]
	\centering
	\caption{Test Coverage Analysis\protect\\Augmenting EvoSuite's test cases with asserts generated by our approach}
	\label{tab:coverage}
	\resizebox{0.8\linewidth}{!}{
\begin{tabular}{lcccccc}
\toprule
\multirow{ 2}{*}{Focal Method} & \multicolumn{2}{c}{EvoSuite} & \multicolumn{2}{c}{EvoSuite + Our Approach} & \multicolumn{2}{c}{Delta Improvement}\\
& Lines & Conditions & Lines & Conditions & Lines & Conditions\\
\midrule
\texttt{toInt(String)} & 21 (5.6\%) & 1 (0.3\%) & 22 (5.9\%) & 2 (0.6\%) & +1 & +1\\
\texttt{toLong(String, long)} & 20 (5.3\%) & 1 (0.3\%) & 21 (5.6\%) & 2 (0.6\%) & +1 & +1\\
\texttt{toFloat(String, float)} & 20 (5.3\%) & 1 (0.3\%) & 21 (5.6\%) & 2 (0.6\%) & +1 & +1\\
\texttt{toDouble(String, double)} & 20 (5.3\%) & 1 (0.3\%) & 23 (6.1\%) & 2 (0.6\%) & +3 & +1\\
\texttt{toByte(String, byte)} & 20 (5.3\%) & 1 (0.3\%) & 21 (5.6\%) & 2 (0.6\%) & +1 & +1\\
\texttt{toShort(String, short)} & 20 (5.3\%) & 1 (0.3\%) & 21 (5.6\%) & 2 (0.6\%) & +1 & +1\\
\texttt{createFloat(String)} & 20 (5.3\%) & 1 (0.3\%) & 21 (5.6\%) & 2 (0.6\%) & +1 & +1\\
\texttt{createDouble(String)} & 20 (5.3\%) & 1 (0.3\%) & 21 (5.6\%) & 2 (0.6\%) & +1 & +1\\
\texttt{createInteger(String)} & 20 (5.3\%) & 1 (0.3\%) & 21 (5.6\%) & 2 (0.6\%) & +1 & +1\\
\texttt{createLong(String)} & 20 (5.3\%) & 1 (0.3\%) & 21 (5.6\%) & 2 (0.6\%) & +1 & +1\\
\texttt{createBigInteger(String)} & 28 (7.5\%) & 8 (2.4\%) & 28 (7.5\%) & 9 (2.7\%) & - & +1\\
\texttt{createBigDecimal(String)} & 22 (5.9\%) & 3 (0.9\%) & - & - & - & -\\
\texttt{min(long[])} & 27 (7.2\%) & 6 (1.8\%) & 27 (7.2\%) & 6 (1.8\%) & - & -\\
\texttt{min(int, int, int)} & 22 (5.9\%) & 2 (0.6\%) & 22 (5.9\%) & 2 (0.6\%) & - & -\\
\texttt{max(float[])} & 28 (7.5\%) & 7 (2.1\%) &  28 (7.5\%) & 7 (2.1\%) & - & -\\
\texttt{max(byte, byte, byte)} & 23 (6.1\%) & 2 (0.6\%) & 23 (6.1\%) & 2 (0.6\%) & - & -\\
\texttt{isDigits(String)} & 20 (5.3\%) & 1 (0.3\%) &  23 (6.1\%) & 5 (1.5\%) & +3 & +4 \\
\texttt{isNumber(String)} & 44 (11.7\%) & 29 (8.6\%) & 45 (12.0\%) & 31 (9.2\%) & +1 & +2 \\
\bottomrule
\end{tabular}}
\end{table*}

\textbf{RQ$_5$: Can our approach be used to improve automatically generated test cases?}
Table \ref{tab:coverage} reports the absolute (and percentage) line and condition coverage at class-level, for each of the 18 public methods considered in the experiment. The table shows the results for the original EvoSuite test cases, those augmented by our model, as well as the delta improvement in the last column.

For 13 out of 18 methods, our approach generated asserts that improved the line and/or condition coverage between 1-3 more lines and 1-4 additional condition coverage. For 4 methods our approach generated correct asserts which did not improve the coverage, while for one method (\ie \texttt{createBigDecimal}) our approach did not generate any correct assert within the top-10 predictions.

Figure \ref{fig:evosuite-asserts} shows all the generated asserts which have been used to augment the EvoSuite test cases, in the same order as the methods reported in table \ref{tab:coverage}. We can notice that the first three asserts invoke the focal method with an actual numerical value, which results in additional test coverage, since the original EvoSuite test case tested the same focal methods with a null or empty string, resulting in the execution of a different branch. The fifth assert, related to the focal method \texttt{toDouble} invokes the focal method using a non-numerical string "foo", and covering three additional lines and one more condition in the focal method, w.r.t. the EvoSuite test case. 

Let us now focus on the four assert statements that did not improve the coverage, corresponding to the focal methods \texttt{min} and \texttt{max}, shown as the sixth to the third from the bottom of figure \ref{fig:evosuite-asserts}. Three of these asserts simply perform additional checks on the return variables used by EvoSuite, namely \texttt{long0, float0, byte0}. These asserts do not invoke the focal method, thus not resulting in additional coverage, but instead focus on testing additional properties of the retun values. Finally, the assert \texttt{assertEquals(4, NumberUtils.min(4, 5, 7))} correctly invokes the focal method and asserts the correct behavior, but executes lines and branches already tested by the original test case (with different values).

Overall, these results show that our approach can be helpful in augmenting existing or automatically generated test cases with additional accurate assert statements. In most of the cases reported in our experiment, we found the asserts to slightly improve the test coverage.

\vspace{-0.0cm}
\begin{center} 
\fbox{
\begin{minipage}[t]{0.97\linewidth}
{\bf Summary for RQ$_5$.} 
Our approach can be used to augment existing test cases, such as those generated by EvoSuite, with additional assert statements. Our experiments show that these asserts can lead to improved test coverage.
\end{minipage}
}
\end{center}

\section{Discussion \& Future Work}
Our experimental analysis showed promising results of our approach in generating accurate assert statements for unit test cases. For our future work, we envision two possible scenarios where we can deploy our model with the goal of improving automation in software testing activities.

\subsection{Supporting Developers in writing Test Cases}
Our approach could be used to support developers in writing test cases more efficiently, by suggesting assert statements while defining the test case. In this scenario, we plan to implement our approach as plugin for an IDE, which is then used by developers while writing code as a code completion tool. Our approach could work side-by-side existing code completion approaches, such as Pythia~\cite{svyatkovskiy2019pythia}. The results of RQ$_3$ clearly shows that our approach is more accurate than standard code completion, leading us to suggest a hybrid approach. In this hybrid approach, a standard code completion tool would perform inference on our model when the developer is writing test cases. 

\subsection{Improving Automated Test Case Generation Tools}
The results of RQ$_5$ show that our approach can be used to augment test cases generated by automated test case generation tools, such as EvoSuite, Randoop, and Agitar. In this scenario, our approach could be integrated within an automated test case generation tool, or used as an external tool which augment and revises assert statements in the newly generated test cases.

\begin{figure}[t]
\vspace{0.4cm}
    \centering
\begin{adjustbox}{width=0.5\textwidth}
\footnotesize
\begin{tabular}{c}
\toprule
Generated Assert\\
\midrule
\begin{minipage}[t]{0.45\textwidth}
\begin{minted}[fontsize=\tiny]{java}
assertEquals(5, NumberUtils.toInt("5"))
assertEquals(1, NumberUtils.toLong("1", 1))
assertEquals(6, NumberUtils.toFloat("6", 6), 0);
assertNotNull(NumberUtils.toDouble("foo", 1.0));
assertEquals(1, NumberUtils.toByte("1",(( byte)(1))));
assertEquals(15, NumberUtils.toShort("15",(( short)(15))));
assertNotNull(NumberUtils.createFloat("1"))
assertNotNull(NumberUtils.createDouble("1"));
assertNotNull(NumberUtils.createInteger("1"));
assertNotNull(NumberUtils.createLong("1"));
assertEquals(BigInteger.valueOf(1), NumberUtils.createBigInteger("1"));
-
assertNotNull(long0);
assertEquals(4, NumberUtils.min(4, 5, 7));
assertTrue(( float0 == 0.0F));
assertTrue(( byte0 == 5));
assertTrue(NumberUtils.isDigits("1"));
assertTrue(NumberUtils.isNumber("1"))
\end{minted}
\end{minipage} \\
\bottomrule
\end{tabular}
\end{adjustbox}
\caption{Assert statements generated by our approach for Lang-1-f\protect\\These asserts are used to augment existing EvoSuite's test cases leading to coverage improvement}
\label{fig:evosuite-asserts}
\end{figure}

\section{Threats to Validity}
Threats to \textit{construct validity} concern the relationship between theory and observation and are mainly related to the measurements we performed. In this context, data leakage could represent a threat to the validity of our study. Data leakage refers to the unintentional and accidental sharing of data between the training and test sets. In our case, the threat arises during the pretraining stage on large amount of source code, where the model may have observed similar code to what found in the finetuning test set. We mitigated this threat by constructing the finetuning process differently from the pretraining, where the code is organized in a dissimilar fashion. Specifically, during the pretraining the test method did not contain the placeholder, and was not adjacent to the focal method. We empirically validated the mitigation of the threat by evaluating the pretrained model (without finetuning) on the test set, in order to observe its performances. The results show that the model was not able to generate correct assert statements, thus confirming the our hypothesis. It is also worth to note that data leakage is avoided within the finetuning dataset (\ie training, validation, test sets) 

\textit{Internal validity} threats concern factors internal to our study
that could influence our results. The performance of our approach
depends on the hyperparameter configuration and pretraining process. We did not perform hyperparameter search since these large models require substantial training time, however, we reuse configurations suggested in the literature. We experiment with different pretraining stages and report the results of our experiments.

Threats to \textit{external validity} concern the generalization of our findings. In our context, the threat arises when comparing our BART Transformer model (with 400M trainable parameters) against the RNN-based model (with 4M trainable parameters) having different capacity and number of parameters. While we note that comparing models with the same number of parameters could yield different results, the authors of ATLAS \cite{watson2020learning} did not observe improvements when increasing the number of layers and units of the Encoder-Decoder architecture. Moreover, we rely on the existing literature comparing Transformer and RNN architectures. 

\section{Related Work}
Our work is related to several existing approaches in the area of automated software testing. In particular, there is a class of approaches that aims at generating tests methods and synthesizing assert statements, such as Evosuite \cite{fraser2011evosuite}, Randoop \cite{pacheco2007randoop}, and Agitar \cite{agitar}. Among these, Evosuite is one of the most popular tools for test generation in Java. It relies on mutation testing in order to generate appropriate assert statements. Specifically, it first introduce mutants within the method under test, then it attempts to generate asserts with the goal of killing the aforementioned mutants. During this process, Evosuite optimizes towards maximizing the number of killed mutants while generating as few asserts as possible. Randoop generates assert statements by relying on user-specified contracts. These statements are then refined using random testing and analyzing execution traces of
the statement it creates. 

The major difference between these works and our approach is the learning component. Specifically, the aforementioned works rely on handcrafted rules or heuristics to generate assert statements, for example via predefined mutations. Instead, we aim to learn from developers' code what are the assert statements that are more effective for the particular context (\ie test case and focal method).

Additionally, recent works have shed light on the importance of generating accurate and complex assert statements to detect real faults in the system \cite{almasi2017industrial, shamshiri2015automated}. In particular, Almasi \etal \cite{almasi2017industrial} shows that, while Evosuite and Randoop were able to uncover several faults in real programs, nearly half of the undetected faults could have been detected with a more appropriate assert statement \cite{almasi2017industrial, watson2020learning}.

These limitations motivated the work from Watson \etal \cite{watson2020learning}, where the authors proposed an RNN-based approach ATLAS which aims at generating meaningful assert statements by learning from developers' code. Inspired by this work, we improve upon it in several substantial ways. First, we employ a different and more advanced deep learning architecture based on transformer models. Second, differently from ATLAS, we take advantage of English and source code semi-supervised pretraining to significantly boost the performances of the models on the assert generation task. Lastly, we investigate qualitative cases and intrinsic metrics as well as the effect of the focal method, which provides additional beneficial context to the model. These contributions culminated into an approach that significantly outperforms the previous work \cite{watson2020learning}, with an 80\% relative improvement in top-1 accuracy.

Our work is related to a broad set of literature on transfer learning~\cite{raffel2019exploring}, unsupervised language model pretraining~\cite{gpt,gpt2}, and denoising pretraining~\cite{bert, roberta, lewis2019bart}. In this paper, we extend these ideas to source code as a language, combining English and source code pretraining modes, fine-tuning on a downstream translation task from the automated software engineering domain.

\section{Conclusion}
In this paper we presented an approach for generating accurate assert statements for unit test cases. The core of our approach is based on a state state-of-the-art transformer model which has been pretrained, in a semi-supervised fashion, on both English and source code corpora. This pretraning process allows to learn the semantics of the natural language and its words as well as the syntax of the source code. The model was then finetuned on the assert generation task, which we represent as a translation task, where the input is the focal method along with the partially generated test case, and the output is the desired assert statement.

The resulting model is able to generate accurate assert statements, with a 62\% top-1 accuracy, matching the exact assert statement originally wrote by the developer. This represents an 80\% relative improvement over the previous RNN-based approach \cite{watson2020learning}.

In our empirical evaluation, we experimented with different pretraining levels, showing the beneficial impact of pretraining on both English and source code in terms of extrinsic and intrinsic metrics. We qualitatively analyzed the assert statements predicted by our model, and identified both common and complex asserts. Interestingly, we found many cases in which the predicted assert statement did not syntactically match the original statement, yet was semantically equivalent and correct. Finally, we empirically demonstrate how our proposed approach can be used to augment existing test cases, such as those generated by EvoSuite, with additional assert statements that lead to test coverage improvements.

We believe that these results are particularly important in terms of the usability and applicability of this approach beyond research, in actual development environments. Practically, developers would obtain accurate and relevant assert statements in one or very few suggestions, allowing them to write complete and robust test cases.


\bibliographystyle{IEEEtran}
\bibliography{main}

\end{document}